\documentclass[manuscript]{aastex}
\newcommand{\science}{Science}

\shorttitle{TTVs analysis in Southern Stars: the case of WASP-28}
\shortauthors{Petrucci et al.}

\usepackage{graphics}
\usepackage{graphicx}
\begin{document}

\title{TTVs analysis in Southern Stars: the case of WASP-28}

\author{R. Petrucci\altaffilmark{1,3,4},
  E. Jofr\'e\altaffilmark{2,3,4}, M. Melita\altaffilmark{1,3},
  M. G\'omez\altaffilmark{2,3}, and
  P. J. D. Mauas\altaffilmark{1,3}}

\altaffiltext{1}{Instituto de Astronom\'{i}a y F\'{i}sica del Espacio (IAFE), Buenos Aires, Argentina.}
\altaffiltext{2}{Observatorio Astron\'{o}mico de C\'{o}rdoba, C\'{o}rdoba, Argentina.}
\altaffiltext{3}{CONICET, Consejo Nacional de Investigaciones Cient\'{i}ficas y T\'{e}cnicas, Argentina.}
\altaffiltext{4}{Visiting Astronomer, Complejo Astron\'{o}mico El Leoncito operated under
agreement between the Consejo Nacional de Investigaciones Cient\'{i}ficas y
T\'{e}cnicas de la Rep\'{u}blica Argentina and the National Universities of La
Plata, C\'{o}rdoba and San Juan.}

\begin{abstract}
We present 4 new transit observations of the exoplanet WASP-28b
observed between August 2011 and October 2013. Employing another
11 transits available in the literature we compute new ephemeris
and redetermine the physical parameters of the star and the exoplanet.
Considering 3 yrs of observations, we find no periodic TTVs or 
long-term variations of the inclination of the orbit, $i$, or the
depth of the transit, $k$, that could be attributable to the presence
of another planetary mass-body in the system. We also study the 
relations between $i$ and $k$ with different factors that characterize
the light-curves. The fits suggest a possible weak correlation between
$k$ with the red noise factor, $\beta$, and the photometric noise rate,
PNR, and a weak anticorrelation between $i$ and PNR, although more 
points are needed to confirm these trends. 
Finally, the kinematic study suggests that WASP-28 is a thin disk star.   
\end{abstract}

\keywords{stars: individual: WASP-28 --- techniques: photometric --- planetary systems}

\section{Introduction}

To date, more than 1400\footnote{Extracted from \textsf{http://exoplanet.eu/catalog/}}
exoplanets have been discovered. Most of them
were mainly detected by the radial velocity \citep{mayor, marcy}
 and the transit \citep{charbo, henry} techniques.
The former method is based on the detection of periodic
 Doppler shifts in the stellar spectral lines due to the
 movement of the star around the star-planet barycenter.
 The second one is based upon the detection of periodic
 variations of the stellar flux produced by the passage
 of a planet in front of the stellar disk. In the
 last years, a new technique became very popular to search for planets:
 the transit timing variations or TTVs \citep{holman, diaz}.
 In a system  composed only by a star and a transiting planet,
 one should expect that the interval between successive minima be constant. 
However, if there is a third body in the system, periodic
 variations of this interval are anticipated. Therefore, it 
 should be possible to detect other planetary-mass bodies
 around stars with an already known transiting planet. The
 less massive the perturber, the smaller the amplitudes of
 the timing variations, and the higher precision minimum 
 central times (with errors of around seconds) are required.
 Many factors have to be taken into account to reach 
 realistic conclusions about TTVs: sampling of the
observations \citep{kipping}, quality of the light curves, presence of red
noise \citep{carter09, barros13, lendl13}, variable atmospheric conditions during the night (for
ground-base observations), etc. In addition, as it has been
shown in several studies \citep[e.g.][]{sou12, nas13}, for
TTV analysis it is of crucial relevance to perform the same
fitting procedure and error treatment in the transits studied.

On the other hand, there is evidence suggesting that the
probability of finding gas giant planets around main-sequence
stars is an increasing function of the stellar metallicity,
reaching only 3\% for stars with subsolar metallicity
\citep{fischer, santos04, santos05}. With a $\mathrm{[Fe/H]} = -0.29\ \mathrm{dex}$,
the F8 star WASP-28 \citep{anderson} seems to fall in the metal-poor long tail of the metallicity
distribution of the planet-host stars. The planet around this star, WASP-28b,
discovered by the SuperWASP program \citep{Pollacco}, is a
Hot Jupiter-like planet ($M_{\mathrm{P}}=0.907\ M_{\mathrm{Jup}}$, $R_{\mathrm{P}}=1.213\ R_{\mathrm{Jup}}$ and $a=0.044\ \mathrm{AU}$)
describing a prograde and aligned orbit. The age of the system ($5^{+3}_{-2}$ Gyr) suggests the possibility that the current orbit of the planet could be reached by planet-disc migration or via tidal interaction although, if the last possibility is assumed, the fact that the planet had not fallen into the star remains unexplained. In this sense, it becomes relevant the study of TTVs to understand the dynamic 
history of the system \citep[e.g.][]{ford14}. In a recent work, \cite{steffen}
 conclude that, in contrast to the presence of transit timing variations in systems
with long-period Jupiters, hot-Neptunes and hot-Earths, the absence of TTVs in Hot-Jupiter 
systems might indicate different planet formation and dynamical evolution theories. 
The authors suggest that the main mechanism responsable for the formation of Hot-Jupiters
could be the planet-planet scattering.

In this work we carried on a complete and homogeneous study of TTVs
and looked for variations in the photometric parameters of WASP-28b. We considered the influence
of many factors related to the quality of the studied light-curves to
explain the high dispersions found in the long-term behaviour of the
parameters.
Finally, we investigated whether the low metallicity of WASP-28 can be
due to its galatic population membership, through the
calculation of its galactic spatial velocities.

This article is organized as follows: in Section \S\ref{sec.obs}  we present our observations and data reduction. In Section \S\ref{sec.par} we describe
the procedure used to fit the light-curves and to derive the photometric and physical parameters. In
Section \S\ref{sec.var} we discuss the long-term variations of the parameters and calculate the relation between them and different factors that characterize the light-curves. In Section \S\ref{sec.eph} we carry out a TTVs study of the system and calculate new ephemeris. In Section \S\ref{sec.uvw} we describe the kinematic properties of WASP-28 and discuss possible scenarios for the formation of its Hot Jupiter in a metal-poor environment. Finally, in Section \S\ref{sec.con} we present the conclusions.

\section{Observations and data reduction}\label{sec.obs}

The 4 transits of WASP-28b were observed between August 2011 and
October 2013. The observations were carried out with the Telescope Horacio
Ghielmetti (THG) located at Cerro Bureck in the Complejo Astron\'omico El
Leoncito (San Juan, Argentina). This telescope is a remotely-operated MEADE-RCX 400 with a 
40-cm primary mirror and Johnson UBVRI filters. The transit corresponding to 2012, July 26 
was observed
using an Apogee Alta U8300 camera with 3326$\times$2504 5$\mu$ pixels, with a scale of 0.32"/pix
and a 19'$\times$14' field of view. The remaining ones were observed employing
an Apogee Alta U16M camera with 4096$\times$4096 9$\mu$ pixels, resulting in a scale of 0.57"/pix
and a 49'$\times$49' field of view. For the transit of 2013 October 26, the images were strongly defocused 
to minimize the dispersion of the resulting transit.
After testing different bin sizes and filters, we decided to use a bin
of 2$\times$2 to achieve a better temporal resolution. We adopted
integration times ranging from 10 to 120 sec depending on seeing, airmass
conditions and the selected filter. In Table \ref{table:1} we present a log of the observations.

Before every observation the computer clock was
automatically synchronized with the GPS time signal, to be certain that the
times of all the images were expressed in Heliocentric Julian Date based on
Coordinated Universal Time ($HJD_{\mathrm{UTC}}$).  
For each night, we took 10 bias and 8 dark frames.
We averaged all the biases and median
combined the bias-corrected darks.
We employed the standard IRAF\footnote{IRAF is
distributed by the National Optical Astronomy Observatories, which
are operated by the Association of Universities for Research in
Astronomy, Inc., under cooperative agreement with the National
Science Foundation.} tasks to perform bias and dark correction. We did not
perform
the flat-fielding correction to the images, since we found that it introduced 
unwanted errors. 

All the instrumental magnitudes were obtained with the
FOTOMCC pipeline \citep{petru13}, which employs 
aperture photometry choosing the optimal annulus
through the growth-curves technique \citep{howell}. Once the instrumental magnitudes were obtained, we carried out 
the differential photometry.

As one of the purposes of this work is to study TTVs, we decided to include
all other transits publicly available. Besides ours, we employed 11 light-curves published in the Exoplanet Transit 
Database (ETD\footnote{\textsf{http://var2.astro.cz/ETD}}). We considered only those
where the transits were clearly visible. 

The smooth trends present in the light-curves are mainly originated by differences between
the spectral types of the comparison and the exoplanet host-star, differential
extinction and, occasionally, by stellar activity .
To remove these trends, for each light-curve we selected the 
before-ingress and after-egress data-points and fitted 
a second order Legendre polynomial. Only in a few cases a first order polynomial was applied.
Then, we removed the fit from all the data (including transit points) and normalized
the out-of-transit (OOT) data to unit.  

\section{Determination of the parameters}\label{sec.par}

\subsection{Photometric parameters}\label{bozomath1}

We derived fundamental stellar parameters and metallicity ($T_{\mathrm{eff}}$, $\log g$,
$\xi$, [Fe/H]) for WASP-28 using HARPS archival spectra\footnote{Based on observations collected at the La Silla Paranal Observatory, ESO (Chile) with the HARPS spectrograph at the 3.6-m telescope, programme ID 085.C-0393(A).}. A
total of 33 individual spectra were co-added to produce a single
spectrum with an average signal-to-noise of around 100:1, suitable for
spectroscopic analysis. 

These four quantities were derived in LTE
using the FUNDPAR\footnote{Available at http://icate-conicet.gob.ar/saffe/fundpar/} fortran code \citep{saffe}. This
program determines fundamental parameters from a list of Fe lines equivalent
widths, using the 2010 version of the MOOG code \citep{sneden},
and calculates Kurucz ATLAS plane-parallel model atmospheres.
Equivalent widths for 27 Fe I and 12 Fe II  weak and isolated lines
were measured automatically using the ARES code 
\citep{sousa}. The resulting values obtained from
the analysis are: $T_{\mathrm{eff}}=6084 \pm 45\ \mathrm{K}$, $\log g= 4.51 \pm 0.03\  \mathrm{cm\  s^{-1}}$,
$\xi=2 \pm 0.12\ \mathrm{km\ s^{-1}}$, $\mathrm{[Fe/H]}=-0.2 \pm 0.07\ \mathrm{dex}$. 
These results agree with the values of the fundamental parameters published by
\citet{anderson}. However, our uncertainties in $T_{\mathrm{eff}}$ and $\log g$
are much smaller than those calculated by \citet{anderson}, probably due to the different method used to compute these parameters.

We calculated theoretical limb-darkening coefficients with the program 
JKTLD\footnote{http://www.astro.keele.ac.uk/~jkt/codes/jktld.html} by bilinear interpolation of the effective temperature and surface
gravity using the tabulations of \citet{claret}, which were built
employing Kurucz ATLAS9 atmospheric models. As these tabulations do not provide theoretical limb-darkening coefficients for the Johnson-R filter, for those transits observed in this band
we adopted the values tabulated for the Cousin-R filter. For the light-curves obtained
with no filter, we took the average
of the values corresponding to the Johnson-V and Cousin-R bands. 

We used the JKTEBOP
code\footnote{http://www.astro.keele.ac.uk/~jkt/codes/jktebop.html} to fit
all the light-curves. This code assumes that the star and the planet  
have  biaxial ellipsoid projections and computes the light-curve considering concentric
circles over each component. 
For each transit, initially we assumed as adjusted parameters
the inclination ($i$), the sum of
the fractional radii ($\Sigma =r_{\mathrm{\star}}+r_{\mathrm{P}}$) and the ratio of the fractional
radii ($k=r_{\mathrm{P}}/r_{\mathrm{\star}}$).
Here, $r_{\mathrm{\star}} = R_{\mathrm{\star}}/a$ and $r_{\mathrm{P}} = R_{\mathrm{P}}/a$ represent the ratios of the absolute radii of the star and the exoplanet, respectively, to the semimajor axis ($a$).
We also took as free parameters the mid-transit time ($T_{0}$) and the scale factor ($l_{0}$). We adopted as initial parameters for the iteration those determined by \citet{anderson}. We calculated $k$ and $\Sigma$ employing the values of $R_{\mathrm{\star}}$, $R_{\mathrm{P}}$ and $a$ computed in the same paper.
As in \citet{sou12} every light-curve was fitted with four different limb darkening laws: linear,         
quadratic, logarithmic and square-root. For each
law, we tried three different possibilities: 1) both coefficients fixed, 2) the linear coefficient fitted
and the nonlinear fixed and, 3) both coefficients fitted. Finally, we
adopted as the best model for a given light-curve the one with minimum $\chi^2$ and 
realistic values for the adjusted parameters.
To get $\chi_{\mathrm{r}}^2=1$, we multiplied the photometric errors by the square-root of the
reduced chi-squared of the fit. Finally, since the Levenberg-Marquardt
optimization algorithm we employed to get the best-fitting model only computes formal errors for the adjusted parameters, we ran two other tasks implemented in JKTEBOP: Monte Carlo simulations (for which we took 10000 iterations) and the residual permutation (RP) algorithm which takes the presence of red noise into account.
We assumed the median value of the empirical data as the final value of each parameter (except for $T_{\mathrm{0}}$, see Section \S\ref{sec.eph}), and  its error was
conservatively adopted as the largest value obtained for both tasks.
In Fig. 1 we show the 15 transits and the best fit to the data and in Table \ref{table:2} we
 list the photometric parameters determined for each light-curve.

To obtain realistic results,
we considered only the values of $i$, $k$ and $\Sigma$ obtained
from the high quality and complete light-curves of our sample.
A way to evaluate the light-curve quality is to calculate the 
photometric noise rate (PNR). In  \citet{fulton} this parameter is 
defined as:
\begin{equation}
      PNR = \frac{rms}{\sqrt{ \Gamma}},
\end{equation}
 where $rms$ is the standard deviation of the
light-curve residuals and $\Gamma$ is the median number of exposures (including
not only the exposure time but also the readout time) per minute.  
In our case, we considered as the best quality light-curves
those with PNR lower than 5 mmag. Then, taking into account only the values of the photometric 
parameters of these complete and high quality
transits (5 points), we computed the values and errors
of $i$, $k$ and $\Sigma$ as the weighted average and the standard deviation of the sample, respectively. 
The final estimations of these parameters are:

\[
      \begin{array}{lp{0.8\linewidth}}
         i=87.92 \pm 0.45     \\
         k=0.127 \pm 0.013                   \\
         \Sigma=0.131 \pm 0.006                   \\
      \end{array}
   \]

Although JKTEBOP gives less certain but still useful results when light-curves 
are incomplete\footnote{John Southworth, private communication.}, to determine
how well the Levenberg-Marquardt optimization algorithm explores the parameter space 
for partial transits, we tested the influence of the initial parameters in the final
results. Therefore, for the 8 incomplete transits of our sample we computed the photometric
parameters adopting as initial values those published by \citet{anderson} $\pm$ 3 times
the error\footnote{For $i$, $k$ and $\Sigma$ we considered as 
errors those calculated through the 5 best light-curves. For $l_{0}$ and the 
limb-darkening coefficients we assumed 0.001 and 0.01, respectively.}, alternatively.
In the first case we named the final parameters $i_{+}$, $k_{+}$ and $\Sigma_{+}$
 and $i_{-}$, $k_{-}$ and $\Sigma_{-}$ in the latter case.
In Table 3, we show the differences between the parameters listed in Table 2 and those
obtained as we explained before. Except for the transits of the epochs 160 and 367, all the
differences remain within the errors. The two outliers correspond
to the poorest-quality light-curves according to the factor PNR (see 
column 8 of Table 2). \textbf{These results indicate that, because the transit
equation is non linear, the Levenberg-Marquardt algorithm can be
trapped in a local minimum (which may not be the global minimum) and hence, might not 
correctly explore the parameter space.}

\subsection{Physical parameters}\label{bozomath}

To determine the physical parameters we employed the JKTABSDIM
code\footnote{http://www.astro.keele.ac.uk/~jkt/codes/jktabsdim.html}.
This code uses standard formulae \citep{sou09}
 to calculate the absolute dimensions
of a system with two components, from the results of radial velocity and light-curve analysis.
As input, it requires the photometric quantities ($i$, $r_{\mathrm{\star}}$,
$r_{\mathrm{P}}$) obtained in the previous section,
the orbital period ($P$) determined from the ephemeris, the stellar velocity amplitude
($K_{\mathrm{\star}}$), for which we adopted the value given by \citet{anderson}, the eccentricity, for which we assumed a
circular orbit ($e=0$), the velocity amplitude of the planet ($K_{\mathrm{P}}$), and the corresponding errors\footnote{For the photometric
parameters we considered the error as
the larger between $\sigma_{\mathrm{+}}$ and $\sigma_{\mathrm{-}}$.}.

The value chosen for $K_{\mathrm{P}}$ was the one which minimizes the figure of merit:  

\begin{equation}
fom = \Bigg[\frac{r^{\mathrm{(obs)}}_{\mathrm{\star}}-(R^{\mathrm{(calc)}}_{\mathrm{\star}}/a)}{\sigma(r^{\mathrm{(obs)}}_{\mathrm{\star}})}\Bigg]^2+\Bigg[\frac{T^{\mathrm{(obs)}}_{\mathrm{eff}}-T^{\mathrm{(calc)}}_{\mathrm{eff}})}{\sigma(T^{\mathrm{(obs)}}_{\mathrm{eff}})}\Bigg]^2,
\end{equation}

\noindent where the predicted radius ($R^{\mathrm{(calc)}}_{\mathrm{\star}}$) and effective
temperature ($T^{\mathrm{(calc)}}_{\mathrm{eff}}$) of the star were determined by
linearly interpolating the calculated stellar mass\footnote{The stellar mass
was computed from Eq. (5) of Southworth (2009).}
 and [Fe/H]\footnote{We adopted the value
 $\mathrm{[Fe/H}]=-0.2\ \mathrm{dex}$, determined in the previous section.} within tabulated
theoretical models. To avoid any dependence of the resulting parameters
with the employed stellar-model we performed this analysis for 3 different 
tabulated models: $Y^2$
\citep{dem04}, Padova \citep{girardi}  and Teramo                    
\citep{pie04}. For each one, we considered
series of isochrones bracketing the lifetime of the star in the 
main-sequence. In this way, it was possible to estimate the age of 
the system. 
We adopted as final value for $K_{\mathrm{P}}$ the average of the
amplitudes given by each model, and for the velocity error we used
the standard deviation. Therefore, by applying this procedure, we obtained $M_{\mathrm{\star}}$, $R_{\mathrm{\star}}$,
$\log g_{\mathrm{\star}}$,	$M_{\mathrm{P}}$,  $R_{\mathrm{P}}$, $a$ and age. 

For completeness, we calculated the exoplanet surface gravity with \citep{sou07}: 

\begin{equation}
g_{\mathrm{P}} = \frac{2\pi}{P}\frac{\sqrt{(1-e^2)} K_{\mathrm{\star}}}{r^2_{\mathrm{P}} \sin(i)},
\end{equation} 

\noindent and the modified equilibrium temperature \citep{sou10} as:
 
\begin{equation}
T'_{\mathrm{eq}} = T_{\mathrm{eff}}\sqrt{\frac{R_{\mathrm{\star}}}{2a}},
\end{equation}   
 
\noindent assuming that the planetary albedo is zero. In Table 4 we present
our results.  The final values for the  physical properties and their errors
were calculated considering the photometric parameters determined in Section \S\ref{bozomath1},
which were computed from the best 5 light-curves.
All these parameters are in good agreement, within errors, with the values reported
in the discovery paper by \citet{anderson}.

\section{Long-term variations of the parameters}\label{sec.var}

Since the presence of a perturber (moon or another planet) could produce 
periodic variations of the depth and/or inclination, in Fig. 2 we show the long-term
behaviour of these parameters. It can be seen that in both cases there are
points which apart more than $\pm$ 1$\sigma$ from the mean value. We computed the Lomb-Scargle 
periodogram \citep{horne} to see if there are any periodicities which could be attributed to the presence 
of a perturber, but we did not find any significant peak.
To discern the origin of these dispersions we studied the correlation between
the photometric parameters and different factors related to the quality
of the light-curves.

\subsection{Incomplete transits}\label{bozomath} 
  
We investigated if the lack of transit points influences
the adjusted parameters.
In Fig. 3 we indicate with empty symbols the incomplete transits and in filled symbols the complete ones. It can be seen in Table \ref{table:5} that the distributions overlap  and the median values are similar within an error of $\pm\sigma$. This means that, even in
the absence of some points in the light-curves, we should expect reliable values for the adjusted parameters i and k.
\textbf{However, this result could be contaminated by other factors (such as white and/or red noises) and most importantly, as we showed in Section \S\ref{bozomath1}, for partial transits the Levenberg-Marquardt algorithm might not correctly explore the parameter space. In this sense,
our conclusion agrees with the one found by recent works \citep{gibson, barros11, barros13} which show that incomplete transits could affect the values and errors of the system parameters obtained from these light-curves. Therefore in the following subsections we exclude partial transits from our analysis.}\\

\subsection{Filter}\label{bozomath}

\citet{sea03} show in Fig 11 how observations of transits with low impact parameters, taken in filters with $\lambda \ge 1\ \mu$  can increase the depth of the transit up to 25$\%$. Since WASP-28 has $b=0.21$ and half of the transits studied in this work were observed with no filter and half in the R-band, we analysed if there is a correlation between the measured depth and the employed filter. In the lower panel of Fig. 3 we show with red circles the transits observed in the R-filter and in blue triangles those observed with no filter (and one in the B-band). 
To determine if both datasets represent the same population we calculated the median and $\pm \sigma$ of the samples (see Table \ref{table:4}) only using complete transits.
We could not make a Kolmogorov-Smirnov test because the number of data in both samples is lower than 10.
By performing a visual inspection and considering the data in Table \ref{table:4}, we concluded that there is no evident trend and both distributions overlap. This can be interpreted as that the band in which the transits were observed has no influence in the final $k$ values. 
In the upper panel of Fig. 3, it can be noticed that the same conclusion is reached regarding the inclination, $i$.\\

\subsection{Red Noise}\label{bozomath}

This is the noise produced by systematic errors due to changes in atmospheric conditions, 
airmass, telescope tracking,
relative flat field errors, or a combination of all these factors. It can also be caused
by the intrinsic variability of the targets. The presence of red
noise in the data implies that adjacent data points in a light-curve
are correlated \citep{pont}. 
Although it is not completely well understood, the existence of this kind
of noise leads to an underestimation of the errors in the adjusted parameters, resulting in an inaccurate
determination of them. Red noise can be quantified with the factor
$\beta=\sigma_{\mathrm{r}}/\sigma_{\mathrm{N}}$, defined by \citet{winn08}. Here, 
$\sigma_{\mathrm{r}}$ is obtained by averaging the residuals into M bins of N
points and calculating the standard deviation of the binned residuals,
and $\sigma_{\mathrm{N}}$ is the expected standard deviation, calculated by: 

\begin{equation}
\sigma_{\mathrm{N}} = \frac{\sigma_{\mathrm{1}}}{\sqrt{N}}\sqrt{\frac{M}{M-1}},
\end{equation} 
where $\sigma_{\mathrm{1}}$ is the standard deviation of the unbinned
residuals. In our case, to estimate the parameter $\beta$, we considered
that the duration of the ingress/egress of
the WASP-28b transits is around 20 minutes, and we averaged the residuals in bins 
of between 10 and 30 minutes. Finally, we used the median value
as the red noise factor corresponding to that light-curve. In column 7 of 
Table \ref{table:2}, we show the values obtained. Since
in the absence of red noise $\beta =1$, we see that in almost all transits 
white noise predominates.

In the lower panel of Fig. 4 we show the variation of depth with 
 $\beta$ for complete transits. The red continuous line represents the best linear
fit to the data obtained through weighted least-squares.  
As it can be seen, there is a weak positive correlation indicating that noisier observations
result in larger depths. Considering all the points, $r=0.381$, but if we
exclude the transit with $\beta=2.2$, $r$ increases to 0.577.  This figure seems to suggest
that observations with large red noise would lead to an overestimation of the 
planetary radius. 

In the upper panel of Fig. 4, we plot the variation of $i$ with the red noise.
Contrary to what occurs with $k$, the correlation between these parameters is almost zero ($r=0.0059$), indicating  that the values of the inclination are not influenced by the presence of red noise.

\subsection{Photon Noise Rate}\label{bozomath}

Fig. 5 shows the variation of 
$i$ and $k$ with PNR for complete transits. In the lower panel
we present $k$ vs PNR.
It can be seen that there is a correlation ($r=0.335$), similar to the one found
for $\beta$, which indicates that
low-quality data can result in overestimations of the 
planetary radius for a given value of the stellar radius. 
On the other hand, the values of $i$ are slightly anticorrelated ($r=-0.269$) with PNR, showing that noisy transits could lead to underestimations of the values of the inclination.

\section{Transit ephemeris and timing}\label{sec.eph}

As the minimum central times are uncorrelated with the rest of the parameters,
we fitted each one of the 15 individual transit light curves considering
$T_{0}$ as the only adjusted parameter. For the fixed parameters we adopted the
final values obtained in Section \S\ref{bozomath1} .
 
We used the \citet{eastman} on-line converter to transform the times of 
all the observations to $BJD_{\mathrm{TDB}}$ (Barycentric Julian Date based on Barycentric 
Dynamical Time). 
We assumed that the $T_{\mathrm{0}}$ have a Gaussian distribution, and therefore we adopted for the
mid-transit times the mean values and the symmetric errors ($\pm\sigma$) given
by the best fit to the light-curves. As we explained in Section \S\ref{sec.par}, the errors considered are the
largest between the estimated by Monte Carlo and by the RP algorithms.

In Table \ref{table:6} we present the mid transit times computed for all the light-curves.
If we include all the 15 points into the analysis, we find a $\chi^2_{\mathrm{r}}=3.27$ which seems to suggest the
possibility of variations in the data. However, as it can be noted in Fig. 6, there are 5 outliers
that strongly deviate from the rest of data. All of these correspond to incomplete
transits (magenta triangles).
\citet{gibson} show that in a large percentage of cases, the mid-times
of partial transits are unrealistic. They attribute this to the fact that a
lack of points in the OOT data affects the symmetry of the light-curve
and hence the central transit times, due to an incorrect normalisation. 
On the other hand, the points of the epochs 159, 358 and 367 correspond to  
transits with high level of red noise
(all of them with $\beta \geq 1$), while the ones corresponding
to the epochs 160, 358 and 367 belong to low quality light-curves. 
 \citet{fulton} show that the inclusion of big outliers
 can lead to false conclusions about the existence of TTVs.
Taking all this into account we decided to exclude all partial
transits from our analysis to sure that they are not biasing the results.
To calculate the best
period ($P$) and the minimum reference time ($T_{\mathrm{minref}}$), we fitted a linear model to the 7 remaining data through weighted least-squares. The new ephemeris obtained are: 
\begin{equation}
T_{\mathrm{0}}(E)=T_{\mathrm{minref}} +  E*P
\end{equation}

\noindent where $P=3.408840 \pm 0.000003\ \mathrm{days}$ and $T_{\mathrm{minref}}=2455290.40551 \pm 0.00102\ \mathrm{BJD_{\mathrm{TDB}}}$. Here $E$ represents the epoch. The uncertainties were obtained from the covariance matrix of the fit and were re-scaled
multiplying them by $\sqrt{\chi^2_{\mathrm{r}}}$. In this case, $\chi^2_{\mathrm{r}}=0.7$, implying that the measurements
agree with a linear ephemeris and there is no evident periodic variations in the O-C diagram. 
As it is shown in Fig. 6, the point corresponding to the epoch 382 is outside the area between the $\pm 1\sigma$ dashed lines. We cannot explain the cause of this anomalous value.  

Since the version of the JKTEBOP code employed in this work (version 28)
does not permit to normalise the out-of-transit data-points and to fit the light-curves 
simultaneously, both processes were carried out separately.
However, recent works \citep{gibson, barros13} have noticed that the measured 
transit-times would correlate with the normalisation function, suggesting that 
the normalisation parameters should be taken into account during the fitting
procedure. 
 
For completeness, we investigated the mass of a possible perturber, considering different orbital configurations. To do that, we assumed an external perturbing body whose semimajor axis is much greater than that of the transiting planet and used the Eq. (2) from \citet{holman}:
\begin{equation}
M_{\mathrm{2}}=\frac{16}{45\pi}M_{\mathrm{\star}}\frac{\Delta t}{P_{\mathrm{1}}}\Bigg(\frac{P_{\mathrm{2}}}{P_{\mathrm{1}}}\Bigg)^2 (1-e_{\mathrm{2}})^3
\end{equation}

\noindent where, $P_{\mathrm{1}}$ and $P_{\mathrm{2}}$ are the orbital periods of WASP-28b and the perturber, respectively, $M_{\mathrm{\star}}$ is the stellar mass, $e_{\mathrm{2}}$ is the eccentricity of the perturber and $\Delta t$ represents the variations in the interval between successive transits. For $P_{\mathrm{1}}$ and $M_{\mathrm{\star}}$, we assumed the values obtained in previous sections and for $\Delta t$ we adopted 3.6 minutes, which is the standard deviation of the 7 points considered to do the TTV analysis. For the perturber, we tested the following values for the eccentricity: 0, 0.001, 0.01, 0.05, 0.1 and 0.25 and semimajor axes from 0.08 to 0.35 AU with step 0.01. In Fig. 7 we showed the results obtained. For clarity, we cut the graph in $M_{\mathrm{2}}=8\ \mathrm{M_{\mathrm{Jup}}}$ and $a_{\mathrm{2}}=0.3\ \mathrm{AU}$. If we consider a very eccentric orbit for the perturber ($e=0.25$) and a semimajor axis almost the double of the semimajor axis of the transiting planet ($a=0.08\ \mathrm{AU}$), our TTVs precision would permit to set the maximum mass of an undetected perturber in more than half of the Saturn mass ($M_{2}=0.21\ \mathrm{M_{\mathrm{Jup}}}$). Under the supposition that the perturbing body has a circular orbit ($e=0$) and ($a=0.08\ \mathrm{AU}$), its mass would be half of the Jupiter mass.

We also investigated the masses of possible external perturbers located in the positions of first-order mean-motion resonances with WASP-28b. We employed the Eq. (33) from \citet{agol}:
\begin{equation}
\Delta t=\frac{P_{\mathrm{1}}}{4.5 j}\frac{M_{2}}{M_{2}+M_{1}}
\end{equation}
\noindent where $j$ is the number of orbits between conjunctions. We calculated the values of $\Delta t$ for the first-order resonances 2:1, 3:2, 4:3 and 5:4, ranging the mass of the perturber from 1 $M_{\mathrm{\earth}}$ to 10 $M_{\mathrm{Jup}}$ with step 0.001. Considering our TTVs dispersion, the results permit to rule out the presence of a perturber with a mass greater than 1.9, 2.8, 3.8 and 4.7 $M_{\mathrm{\earth}}$ at the 2:1, 3:2, 4:3 and 5:4 resonances with WASP-28b. It is important to notice that this is just a first approximation and a more rigorous analysis of the resonances, employing  equations of motion, should be done in order to obtain more precise results.

  \section{Kinematic properties of WASP-28}\label{sec.uvw}

In order to establish the Galaxy population membership of the metal-poor star WASP-28, we computed its kinematic properties. The galactic-velocity components ($U$,$V$,$W$) and their errors were calculated following the methodology described in Sections III and IV of \citet{johnson}.
We assumed the same directions for $V$ and $W$, but for the $U$-component we supposed  the opposite direction to the one adopted in that paper. To do the calculation, we used as initial quantities: $RA (2000)= 353.616$,  $DEC (2000)=-01.580$, $pmRA=22.5 \pm 1.3\ \mathrm{mas\ yr^{-1}}$, $pmDEC= 7.8 \pm 1.3\ \mathrm{mas\ yr^{-1}}$ , $RV = 24.33\ \pm\ 0.02\ \mathrm{km\ s^{-1}}$ and $\pi=2.439 \pm 0.014\ \mathrm{mas}$. The values of $pmRA$ and $pmDEC$ were taken from \citet{zac}, the radial velocity was calculated from the same HARPS spectra employed to determine the photometric parameters, whereas the rest of the parameters were taken from \citet{anderson}. The parallax ($\pi$) was calculated from the value of distance published in that paper. 
The final space-velocity components were derived relative to the Local Standard of Rest assuming a solar motion of $(U,V,W)_{\mathrm{\sun}} = (-10.00,+5.25,+7.17)$ \citep{dehnen}. The resulting values for WASP-28 are: $U_{\mathrm{LSR}}=34.56\ \mathrm{km\ s^{-1}}$,$V_{\mathrm{LSR}}=12.84\ \mathrm{km\ s^{-1}}$,$W_{\mathrm{LSR}}=-19.12\ \mathrm{km\ s^{-1}}$.
Based on these velocities we investigated the Galactic population membership of WASP-28 considering the probabilities that the star belongs to the thick or thin disk or the halo ($p_{\mathrm{thin}}$, $p_{\mathrm{thick}}$, $p_{\mathrm{halo}}$). To compute these probabilities, we employed the Eq. (1) and (2) of \citet{reddy}
obtaining the following values: $p_{\mathrm{thin}}=0.985$, $p_{\mathrm{thick}}=0.014$, $p_{\mathrm{halo}}=3.27 \times 10^{-5}$, which suggest that WASP-28 is a thin disk star.
Therefore, this star might have been formed in some of the metal-poor clouds of the local neighborhood.

In the last years several works \citep{ghezzi10b, johnson10, johnson12, maldonado13} have discussed the scenario for the formation of planets around metal-poor stars (or planets in the low metallicity tail of the planet-metallicity correlation), like WASP-28. In the core accretion model \citep{pollack} a threshold density of solid material in the protoplanetary disk is necessary for a rapid growth of planetesimals. Once a massive enough core is formed, it can accrete an atmosphere forming a gas-giant planet which can migrate close to the star, before the gas dissipates. This model is strongly dependent on the metallicity of the cloud \citep{matsuo}. This means that the larger the metallicity the faster the formation of the giant planet, giving it enough time to migrate to distances less than $0.1\ \mathrm{AU}$. But, if the metallicity of the disk is low, the growth of planetesimals is slower. Therefore, when the giant planet is completly formed, much of the gas near the star is already dissipated, and the recently formed planet cannot migrate too close to the star. Therefore, protoplanetary disks with low metallicity would have a scarce probability of forming Hot Jupiters. However, \citet{natta} showed that high-mass stars would have massive protoplanetary disks. Recently, \citet{alibert} and \citet{mordasini} showed that giant planet formation can take place in protoplanetary disks with low metallicity and high mass. In this scenario, the required limit of metal content to form a Jupiter-like planet would be lower for disks around high-mass stars than for disks surrounding less massive stars. \citet{johnson10} also showed that giant-planet frequency is an increasing function of not only the metallicity but also the stellar mass, and hence of the mass of the disk. In this scenario, the stellar mass would compensate the protoplanetary-disk low metallicity, allowing the formation of Jupiter-like planets \citep[and references therein]{kennedy, ghezzi, johnson10}. As it was noted by \citet{ida} and \citet{lau} if the disk mass increases, then the surface density of the protoplanetary disk increases, favoring the formation of gas giant planets in the core accretion model. 

Another possibility is the gravitational-instability model of \citet{boss06}.  \citet{boss97, boss02} proposed this mechanism to explain the formation of giant planets before the dissipation of gas in the disk. In this scenario, if the protoplanetary disk is massive enough, it could break up in dense fragments. Tipically, in several hundreds of years these fragments would contract to form  giant proto-planets. In this situation, the planets form quickly, before the gas depletion. Contrary to the core-accretion theory, this model is almost independent on [Fe/H]. Furthermore, \citet{cai} and \citet{meru} pointed out that the efficiency of the formation of giant planets through disk instability decreases as the metallicity increases. 

Considering that WASP-28 is an intermediate-mass star, we can assume that its protoplanetary disk was of intermediate mass. In this scenario, both models could explain the origin of WASP-28b. On one hand, if the planet was formed through the core accretion model, the low-metallicity disk could have been compensated by the required density of solid material provided by the mass in the disk. If the planet was formed by Boss' model, the abundant mass of the protoplanetary disk would have been enough to allow the fragmentation and consecutive collapse of the resulting proto-planets. 

\section{Conclusions}\label{sec.con}

In this work we presented observations of 4 new transits of the
system WASP-28, observed  between August 2011 and October
2013. Additionally, we also studied 11 transits reported by other
observers, which were downloaded from the Exoplanet Transit Database.

We performed an homogeneous study of the 15 transits, and obtained
new ephemeris and redetermined the parameters of the system. To look
for evidence of another planetary-mass body present in the system,
we analysed the long-term variations of $i$ and $R_{P}/R_{\star}$,
finding large scattering in some points but without evidence of a
periodic pattern. We found a weak anticorrelation between the noise of
the observations (measured by PNR) and the $i$ measured in the
transit, indicating that the inclination measured in noisy transits
could be underestimated. For $R_{P}/R_{\star}$, we found a possible 
correlation with red noise and PNR, which would suggest that noisy
light-curves, would lead to overestimate the planetary radius for a
given value of $R_{\star}$. However it is important to caution that
these relations are based on a small number of points ($N=7$) and a
statistically significant sample  is needed to confirm these trends.

We performed the first TTV analysis of this system. We found that the
$O-C$ are well fitted by a linear ephemeris, with the exception of several
outliers, which correspond to incomplete
transits and very noisy observations. Therefore, we found no evidence of the presence of another body
in the system.

On the other hand, the dynamical analysis showed that for our TTVs precision
the maximum mass that a perturber could have is $M_{2}=0.21$ $M_{Jup}$, considering an eccentric
orbit. However, if the orbit of the perturber is circular the maximum
mass should be of $M_{2}=0.51$ $M_{Jup}$. In the case of
2:1, 3:2, 4:3 and 5:4 first-order resonances we found that our data permit
to exclude a external perturber with mass greater than 1.9, 2.8, 3.8 and 
4.7 $M_{\mathrm{\earth}}$ respectively.

Finally, we performed the first study of the kinematic properties of
WASP-28. We measured the components of the galactic velocity
($U_{LSR}$, $V_{LSR}$, $W_{LSR}$) and found that there is a
probability of $0.94$ that the star belongs to the thin disk.

\section*{Acknowledgments}

We thank Mart\'in Schwartz for his valuable work in the maintenance and operation of the THG, Pablo Perna and the Casleo staff for technical support, and CONICET for funding this research. R.P. also thanks John Southworth for useful comments and Andrea Buccino for her help with the references. R.P. and E.J. are also grateful to A. Sabella for the motivation. We also thank the anonymous referee for his/her useful comments and suggestions, mainly those corresponding to the Sections 3.1 and 4.1, which really helped to improve the quality of this paper. This research has made use of the SIMBAD database, operated at CDS, Strasbourg, France and NASA's Astrophysics Data System.

\bibliographystyle{mn2e}

\clearpage

\begin{figure*}
   \centering
   \includegraphics[width=.8\textwidth]{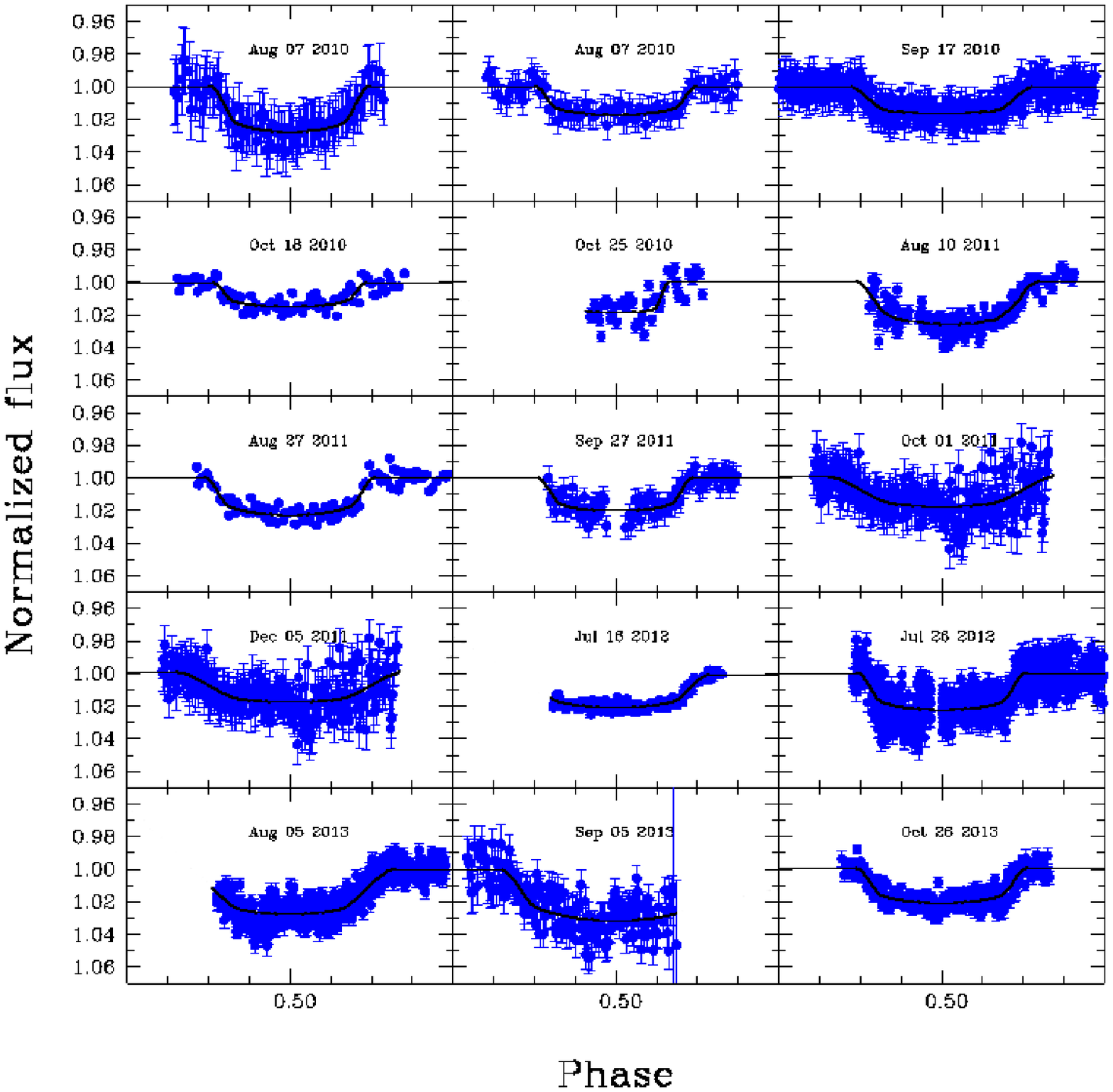}
      \caption{Transits analysed in this work. The photometric data and their error bars are marked in blue. Black solid lines represent the best fit to the data. We also indicate the date when the transits were observed.}
         \label{FigVibStab}
   \end{figure*}

\begin{figure}
   \centering
   \includegraphics[width=\hsize]{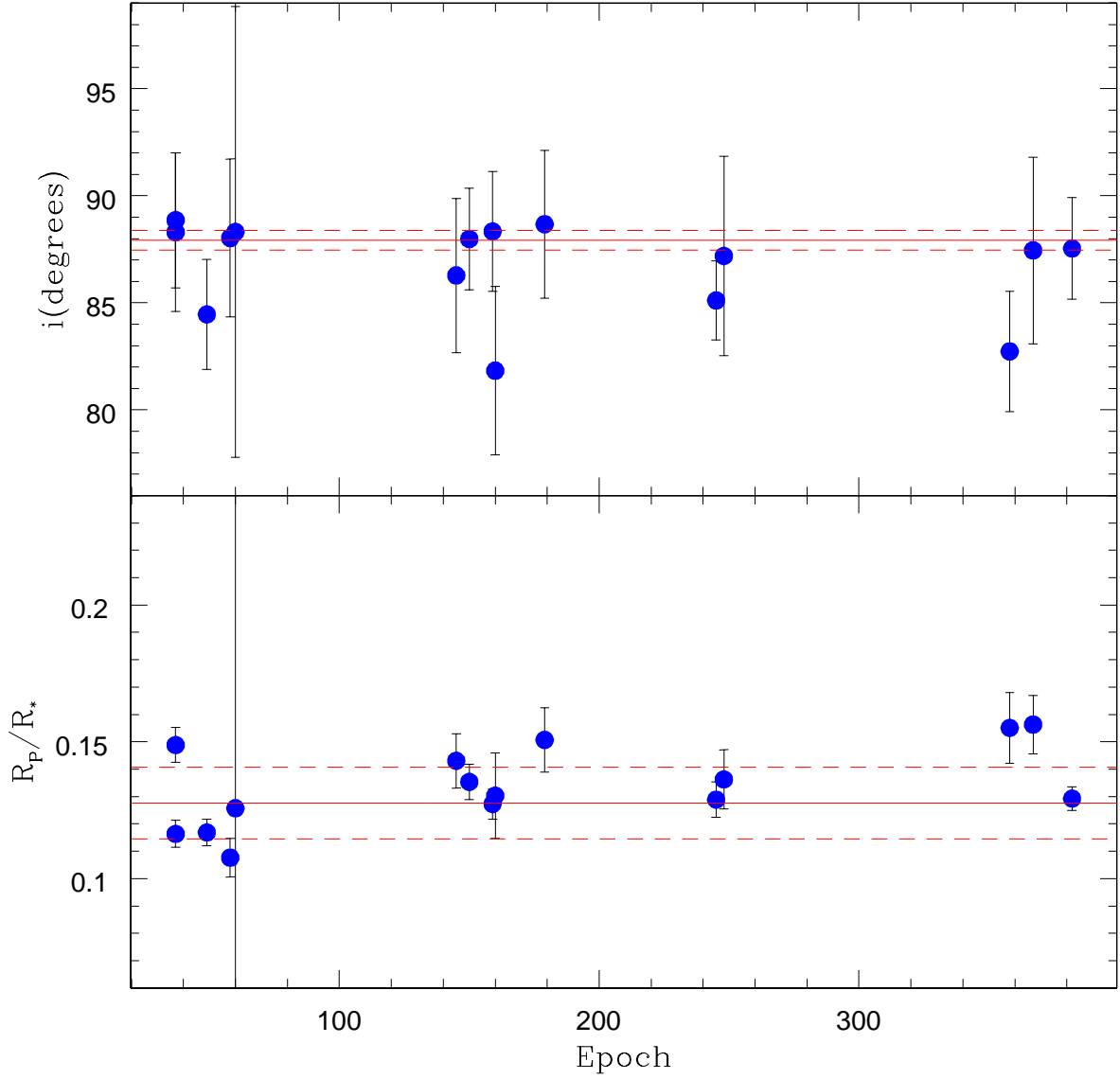}
      \caption{Long-term variation of $i$ (upper panel) and $k$ (lower panel). 
      Continuous lines
      represent the weighted averages calculated in Section \S\ref{bozomath1} and dashed lines indicate 
      $\pm$ 1$\sigma$. Error bars are also shown. }
         \label{FigVibStab}
   \end{figure}

\begin{figure}
   \centering
   \includegraphics[width=\hsize]{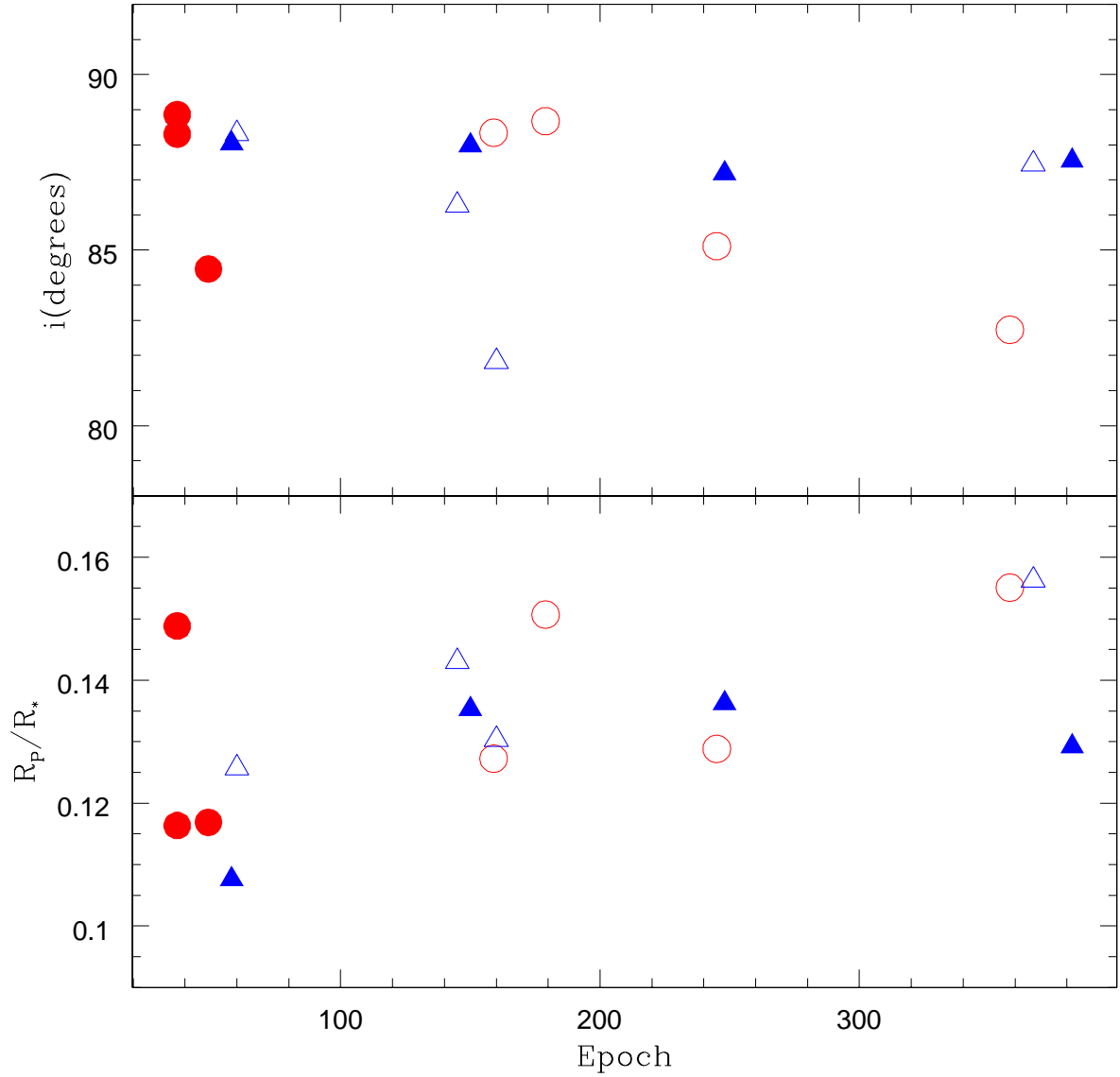}
      \caption{Long-term variation of $i$ (upper panel) and $k$ (lower panel). 
      The filled and empty symbols indicate the complete and incomplete transits, respectively. Red circles correspond to the transits taken in the red filters and blue triangles to those observed in the blue filters (B or no filter).}
         \label{FigVibStab}
   \end{figure}

   \begin{figure}
   \centering
   \includegraphics[width=\hsize]{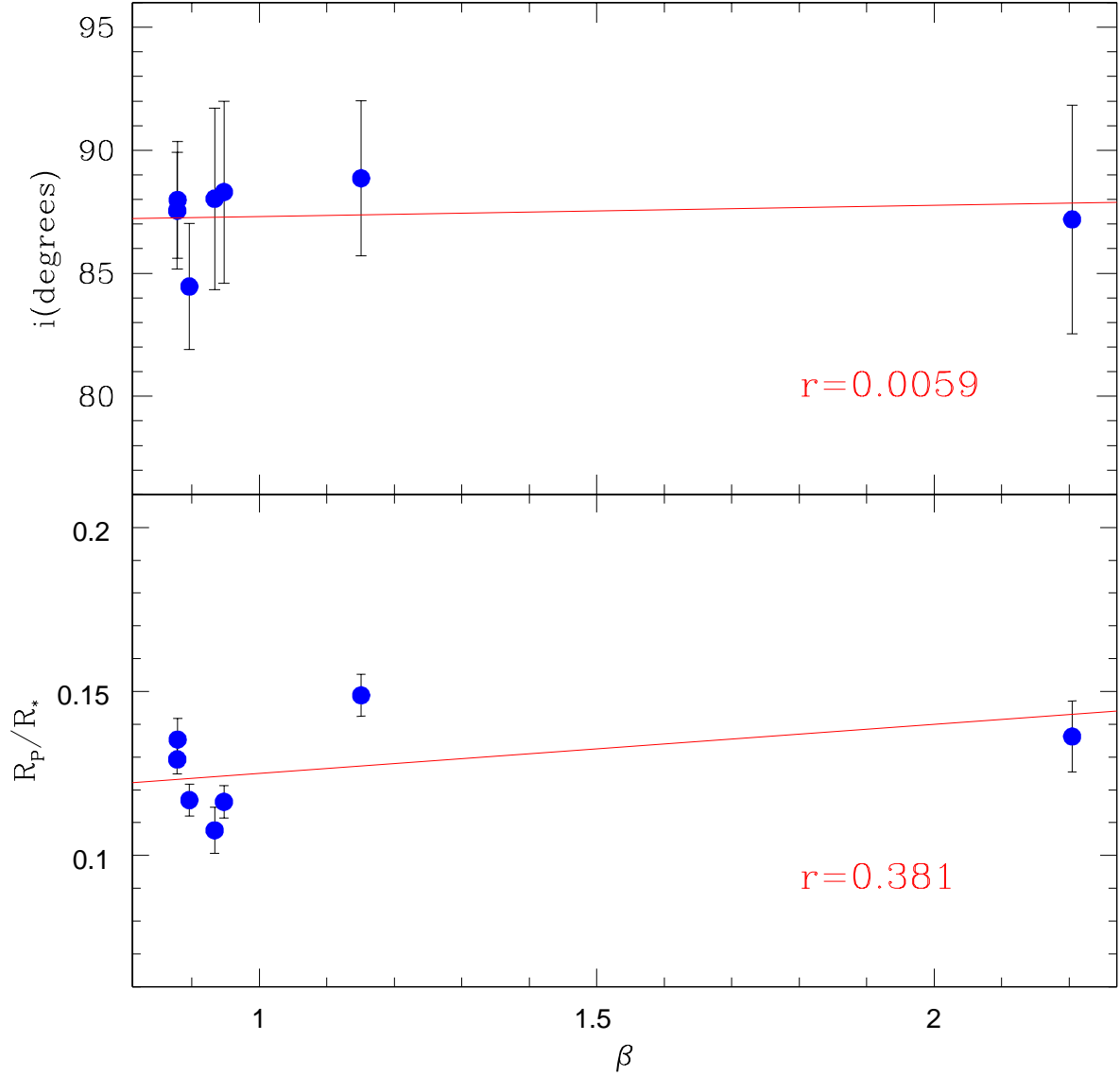}
   \caption{Variation of $i$ and
    $k$ with the red noise factor, $\beta$. The red continuous lines are the best linear
    fit to the data obtained through weighted least-squares. The correlation coefficients $r$
    are also shown. In the lower panel the $r=0.577$ is obtained excluding the point with the largest
    $\beta$. If it is included, the correlation coefficient is $r=0.381$. Error bars are also shown.}
              \label{FigGam}%
    \end{figure}

%                                     Two column figure (place early!)
%______________________________________________ Gamma_1 (lg rho, lg e)
   \begin{figure}
   \centering
   \includegraphics[width=\hsize]{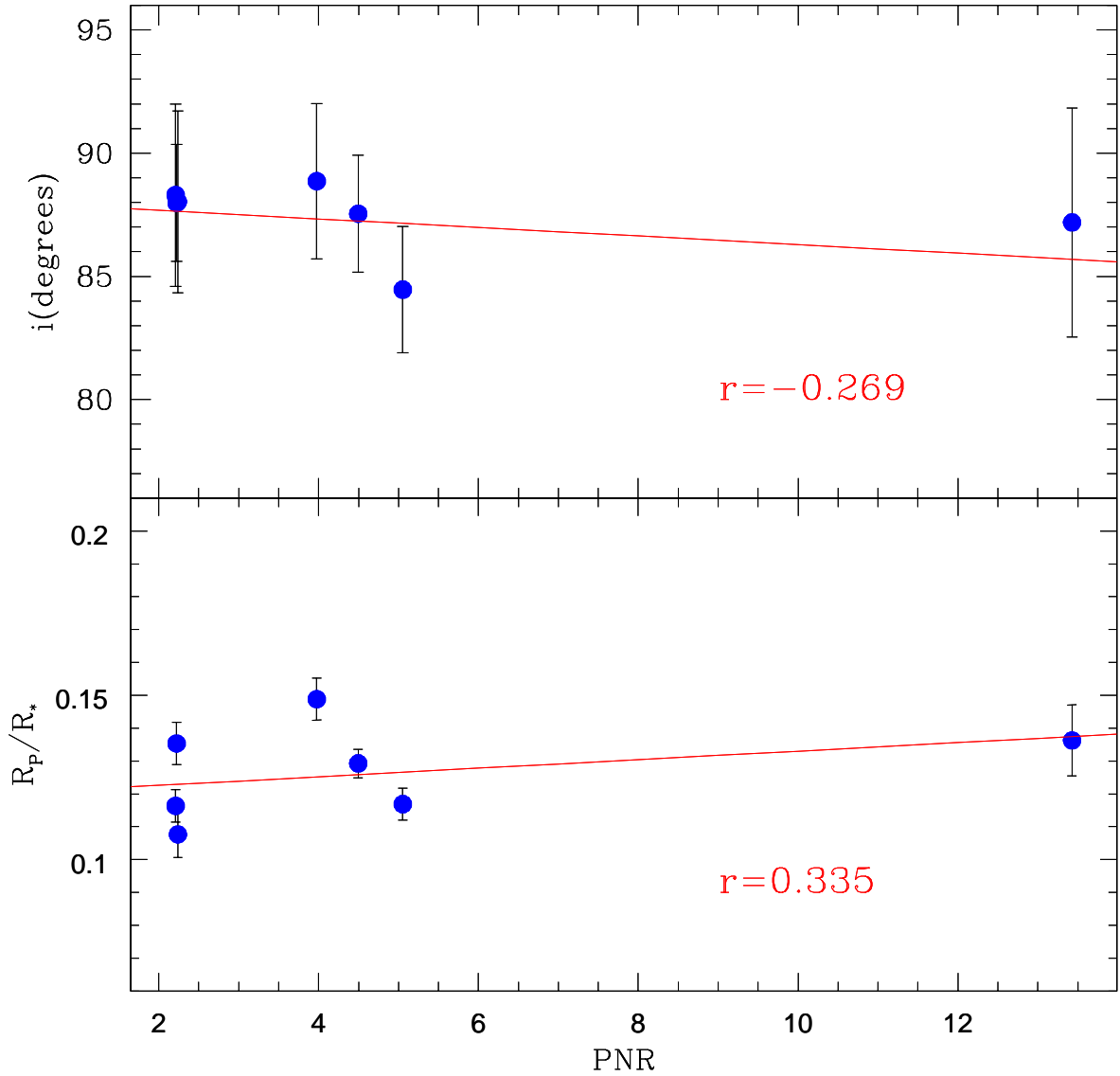}
   \caption{Variation of $i$ and
    $k$ with PNR. The red continuous lines are the best linear
    fit to the data obtained through weighted least-squares. Error bars are also shown.}
              \label{FigGam1}%
    \end{figure}

\begin{figure}
   \centering
   \includegraphics[width=\hsize]{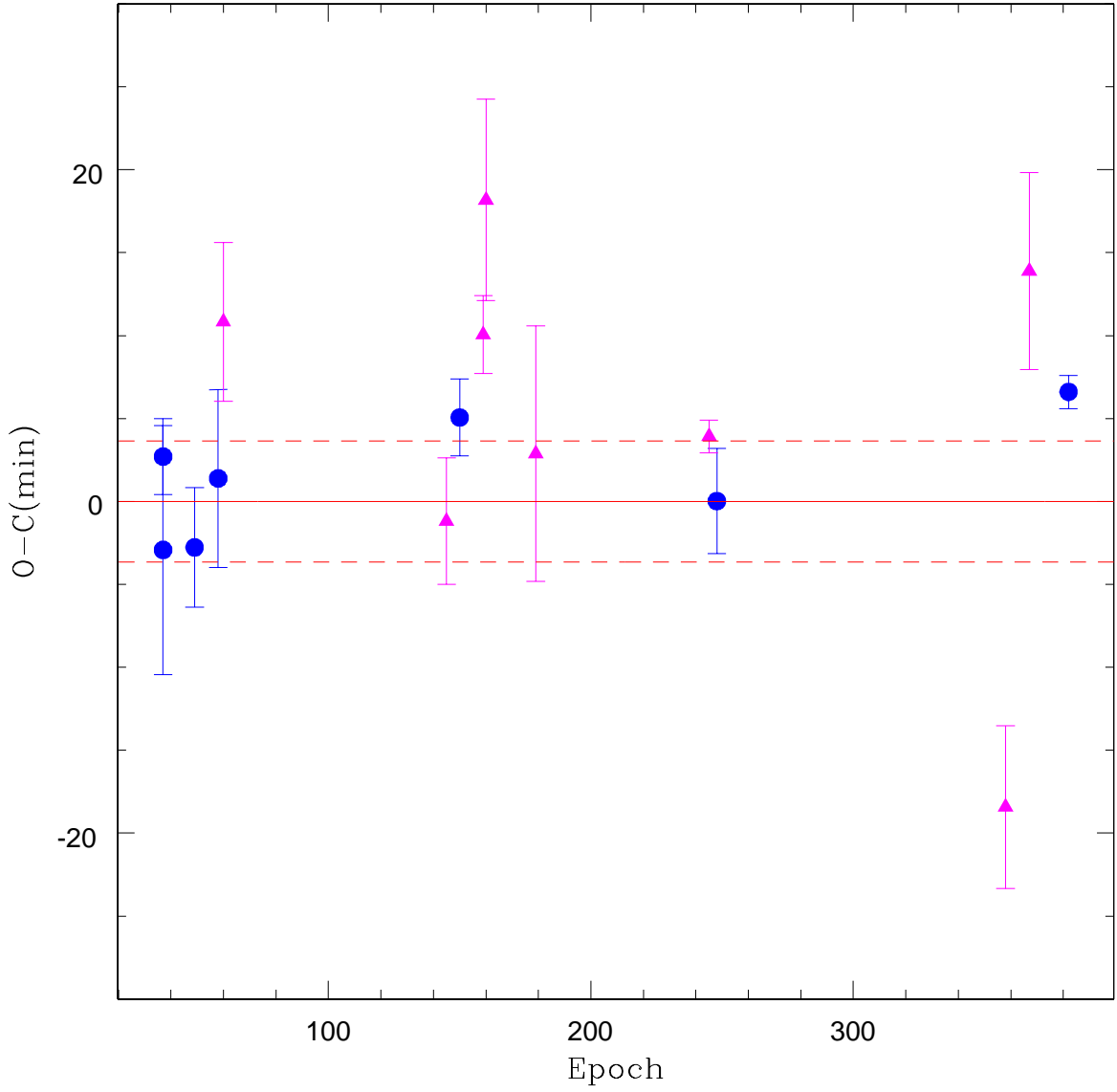}
      \caption{O-C diagram for transit timing of WASP-28b. Magenta triangles mark 
      the points excluded from our calculation of the new ephemeris. Dashed lines
      indicate $\pm 1\sigma$ considering only the blue circles.}
         \label{FigVibStab}
   \end{figure}

 \begin{figure}
   \centering
   \includegraphics[width=\hsize]{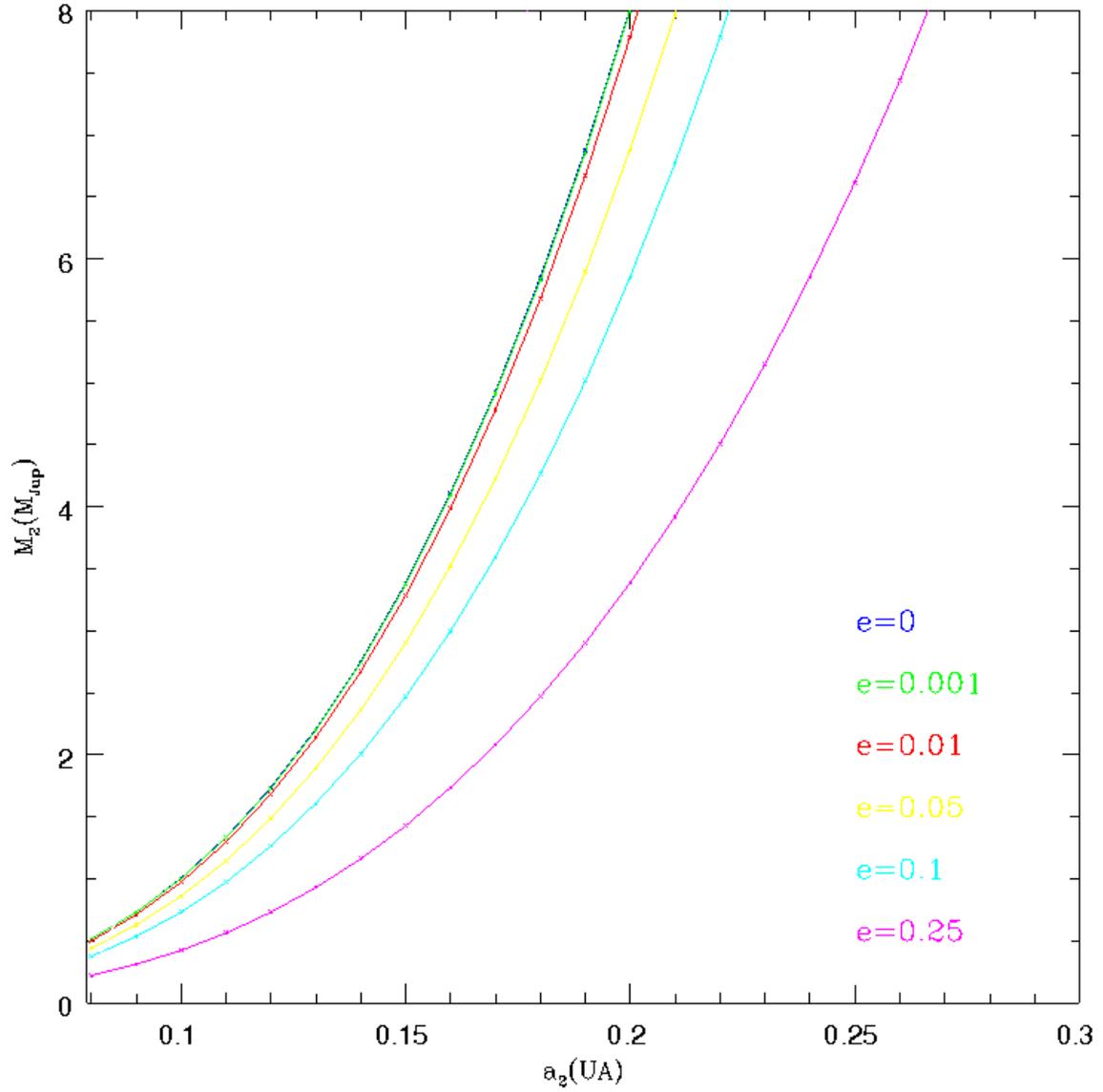}
      \caption{Semimajor axis vs mass of a possible perturbing body. Each color indicates a different eccentricity.}
         \label{FigVibStab1}
   \end{figure}
 
\clearpage

\begin{deluxetable}{lcccccc}
\tabletypesize{\scriptsize}
\tablecaption{Log of our observations. $N_{obs}$ is the number of data-points and $\sigma$
 is the standard deviation of the out-of-transit data-points.
\label{table:1}}
\tablewidth{0pt}
\tablehead{
\colhead{Date} & \colhead{Camera} & \colhead{Filter} & \colhead{Bin-size} &
\colhead{Expsoure-Time (s)} & \colhead{$N_{obs}$} & \colhead{$\sigma$ (mag)} 
}
\startdata
Aug 27 2011 & U16M & no filter & 1x1  & 120 & 105 & 0.0042 \\
Jul 26 2012 & U8300 & no filter & 4x4   &  10 & 711 & 0.0078\\
Aug 5 2013 & U16M & R & 2x2   & 30 & 407 & 0.0049 \\
Oct 26 2013 & U16M & no filter & 2x2   & 50  &  231 & 0.004\\
\enddata
%% Text for table notes should follow after the \enddata but before
%% the \end{deluxetable}. Make sure there is at least one \tablenotemark
%% in the table for each \tablenotetext.
\end{deluxetable}

\clearpage

\begin{deluxetable}{l c c c c c c c c c}
\tabletypesize{\scriptsize}
\tablecaption{Characteristics of the Light Curves and Photometric Parameters Obtained
\label{table:2}}
\tablewidth{0pt}
\tablehead{
\colhead{Date} & \colhead{Epoch} & \colhead{$i$} & \colhead{$k$} & \colhead{$\Sigma$} & \colhead{Filter} & \colhead{$\beta$} &
\colhead{PNR (mmag)} & \colhead{Complete?} & \colhead{Reference} 
}
\startdata
Aug 07 2010$^{*}$	&	37	&	$88.86^{+1.1}_{-3.15}$	&	$0.1488^{+0.0059}_{-0.0064}$	&	$0.1244^{+0.0208}_{-0.0064}$	&	R	&	1.1508	&	3.977	&	Yes	& 1\\
Aug 07 2010$^{*}$	&	37	&	$88.3^{+1.45}_{-3.7}$	&	$0.1164^{+0.0048}_{-0.0049}$	&	$0.1255^{+0.0220}_{-0.0066}$	&	R	&	0.9475	&	2.211	&	Yes	& 2\\
Sep 17 2010	&	49	&	$84.46^{+2.56}_{-2.51}$	&	$0.1169^{+0.0032}_{-0.0048}$	&	$0.1681^{+0.0304}_{-0.0273}$	&	R	&	0.8963	&	5.053	&	Yes	& 3\\
Oct 18 2010$^{*}$	&	58	&	$88.03^{+1.9}_{-3.69}$ 	&	$0.1076^{+0.0070}_{-0.0051}$	&	$0.1215^{+0.0367}_{-0.0081}$	&	no filter	&	0.9337	&	2.239	&	Yes	& 4 \\
Oct 25 2010	&	60	&	$88.31^{+1.65}_{-10.52}$	&	$0.1257^{+0.0140}_{-0.1248}$	&	$0.0901^{+0.0447}_{-0.0231}$	&	no filter	&	0.8813	&	3.949	&	No	& 4 \\
Aug 10 2011	&	145	&	$86.28^{+3.6}_{-1.89}$	&	$0.1431^{+0.0100}_{-0.0062}$	&	$0.1439^{+0.0296}_{-0.0156}$	&	B	&	0.9776	&	4.362	&	No	& 5\\
Aug 27 2011$^{*}$	&	150	&	$87.98^{+1.86}_{-2.37}$ 	&	$0.1353^{+0.0044}_{-0.0064}$	&	$0.1357^{+0.0205}_{-0.0069}$	&	no filter	&	0.8789	&	2.223	&	Yes	& 6\\
Sep 27 2011	&	159	&	$88.34^{+1.21}_{-2.8}$	 &	$0.1272^{+0.0056}_{-0.0050}$	&	$0.1231^{+0.0227}_{-0.0079}$	&	R	&	1.2533	&	3.191	&	No	&  7\\
Oct 1 2011	&	160	&	$81.83^{+2.27}_{-3.93}$	&	$0.1303^{+0.0156}_{-0.0124}$	&	$0.2122^{+0.0697}_{-0.0348}$	&	no filter	&	1.0963	&	8.906	&	No	& 8\\
Dec 5 2011	&	179	&	$88.67^{+1.26}_{-3.46}$ 	&	$0.1507^{+0.0118}_{-0.0068}$	&	$0.1275^{+0.0224}_{-0.0075}$	&	R	&	1.4977	&	5.509	&	No	&  8\\
Jul 16 2012	&	245	&	$85.11^{+1.84}_{-1.24}$ 	&	$0.1288^{+0.0044}_{-0.0065}$	&	$0.1636^{+0.0157}_{-0.0165}$	&	R	&	0.7990	&	1.818	&	No	& 9 \\
Jul 26 2012	&	248	&	$87.19^{+2.57}_{-4.65}$	&	$0.1363^{+0.0108}_{-0.0107}$	&	$0.1402^{+0.0366}_{-0.0148}$	&	no filter	&	2.2044	&	13.428	&	Yes	& 6\\
Aug 5 2013	&	358	&	$82.73^{+2.26}_{-2.81}$ 	&	$0.1551^{+0.0129}_{-0.0063}$	&	$0.2004^{+0.0513}_{-0.0300}$	&	R	&	1.7519	&	7.727	&	No	& 6\\
Sep 5 2013	&	367	&	$87.45^{+2.17}_{-4.36}$	&	$0.1563^{+0.0106}_{-0.0071}$	&	$0.1879^{+0.0533}_{-0.0273}$	&	no filter	&	1.3380	&	8.518	&	No	& 10\\
Oct 26 2013$^{*}$	&	382	&	$87.54^{+2.37}_{-1.69}$	&	$0.1293^{+0.0044}_{-0.0035}$	&	$0.1369^{+0.0155}_{-0.0089}$	&	no filter	&	0.8783	&	4.495	&	Yes	& 6\\
\enddata

$^{*}${Transits used to calculate the final values of $i$, $k$ and $\Sigma$.}
Columns 3-5: Values of the derived photometric parameters and their errors. Column 7: Median value for the red noise. Column 8: Photon noise rate.
References: (1) Lorenz E. R. (TRESCA); (2) Naves R. (TRESCA); (3) Saral G. (TRESCA); (4) Curtis I. (TRESCA); (5) Makely N., Pree C. D. (TRESCA); (6) This work; (7) Gillier Ch. (TRESCA); (8) Shadic S. (TRESCA); (9) Sauer T. (TRESCA); (10) Pouzenc C. (TRESCA).
\end{deluxetable}

\clearpage

\begin{table}
\caption{Differences between $i$, $k$ and $\Sigma$ and the values computed as it is explained in the text, for partial transits}             
\label{table:3}      
\centering          
\begin{tabular}{lccc}     % 7 columns 
\hline\hline       
Epoch	&	$\Delta i_{+}$	&	$\Delta k_{+}$	&	$\Delta \Sigma_{+}$\\
\hline                    
60	&	0.3	&	0.000384	&	0.0009	\\
145	&	0.43	&	-0.001754	&	-0.0023	\\
159	&	0.27	&	-0.001447	&	-0.0021	\\
160	&	5.54	&	-0.018349	&	-0.0645	\\
179	&	-0.24	&	0.00213	&	-0.0019	\\
245	&	-0.16	&	0.000861	&	0.0029	\\
358	&	0.02	&	-0.00029	&	0.0001	\\
367	&	0.12	&	-0.000478	&	-0.0002	\\
\hline\hline							
	&	$\Delta i_{-}$	&	$\Delta k_{-}$	&	$\Delta \Sigma_{-}$\\
\hline    
60	&	-0.01	&	0.002184	&	-0.0009	\\
145	&	0.4	&	-5.4E-05	&	-0.0048	\\
159	&	-0.13	&	0.000753	&	-0.0012	\\
160	&	-4.9	&	0.030651	&	0.0854	\\
179	&	0.21	&	0.00413	&	-0.0039	\\
245	&	-0.47	&	0.001861	&	0.0056	\\
358	&	0	&	0.00021	&	-0.0005	\\
367	&	-7.48	&	0.023822	&	0.0782	\\
\hline                  
\end{tabular}
\end{table}

\begin{table}
\caption{Physical Properties of the Star and the Exoplanet}             
\label{table:4}                
\begin{tabular}{lc}    
\hline\hline
Parameter & Value \\       
\hline
Stellar Mass $M_{\mathrm{\star}}\ (M_{\mathrm{\sun}})$ & 1.011 $\pm$ 0.028\\
Stellar Radius $R_{\mathrm{\star}}\ (R_{\mathrm{\sun}})$ & 1.123 $\pm$ 0.052\\
Stellar gravity $\log g_{\mathrm{\star}}\ (\mathrm{cgs})$ & 4.342 $\pm$ 0.040\\
Planet Mass $M_{\mathrm{P}}\ (M_{\mathrm{Jup}})$ & 0.899 $\pm$ 0.035\\
Planet Radius  $R_{\mathrm{P}}\ (R_{\mathrm{Jup}})$ & 1.354 $\pm$ 0.166 \\
Planet equilibrium temperature $T_{\mathrm{eq}}\ (\mathrm{K})$ & 1473 $\pm$	30\\
Planet surface gravity $\log g_{P}\ (\mathrm{cgs})$ & 3.083 $\pm$ 0.091\\
Semimajor axis $a(UA)$ & 0.0445 $\pm$ 0.0004\\
Age (Gyr) & 4.2 $\pm$ 1.0\\

\hline                  
\end{tabular}
\end{table}

\clearpage

\begin{table}
\caption{Median value and standard deviation for $i$ and $k$ considering complete or incomplete transits}             
\label{table:5}      
\centering                          
\begin{tabular}{c c c c c}        
\hline\hline                 
Transit & $i_{\mathrm{median}}$ & $\sigma_{\mathrm{i}}$ & $k_{\mathrm{median}}$ & $\sigma_{\mathrm{k}}$ \\ 
\hline                       
\hline
Complete	&	87.98 	& 1.43 & 0.129 & 0.014	\\
Incomplete	&	86.86	&	2.64 & 0.136	 & 0.013	\\
\hline                                   
\end{tabular}
\end{table}   

\begin{table}
\caption{Median value and standard deviation for $i$ and $k$ considering red or blue filters.}            
\label{table:4}      
\centering                          
\begin{tabular}{c c c c c}        
\hline\hline                 
Filter & $i_{\mathrm{median}}$ & $\sigma_{\mathrm{i}}$ & $k_{\mathrm{median}}$ & $\sigma_{\mathrm{k}}$ \\ 
\hline                       
\hline
Red & 88.3 & 2.39 & 0.116 & 0.018\\				
Blue & 87.76 & 0.39 & 0.132 & 0.013\\				
\hline                                   %inserts single line
\end{tabular}
\end{table}

\begin{table}
\caption{Mid-transit times}             
\label{table:6}      
\centering                          
\begin{tabular}{c c c c}        
\hline\hline                 
Epoch & $T_{\mathrm{0}}\ (BJD_{\mathrm{TDB}})$ & $e_{\mathrm{T_{\mathrm{0}}}}\ (BJD_{\mathrm{TDB}})$ & Reference \\  
\hline                        
37	&	2455416.529869	&	0.005214	 &	1 \\
37	&	2455416.533777	&	0.001582	 &	2 \\
49	&	2455457.435933	&	0.002501	 &	3 \\
58	&	2455488.118293	&	0.003719	 &	4 \\
60	&	2455494.942504	&	0.003327	 &	4 \\
145	&	2455784.684716	&	0.002644	 &	5 \\
150	&	2455801.733205	&	0.00161 &	6 \\
159	&	2455832.416153	&	0.001629	 &	7 \\
160	&	2455835.830611	&	0.004218	 &	8 \\
179	&	2455900.587763	&	0.005352	 &	8 \\
245	&	2456125.571263	&	0.000681	 &	9 \\
248	&	2456135.795042	&	0.002203	 &	6 \\
358	&	2456510.753533	&	0.003411	 &	6 \\
367	&	2456541.455445	&	0.004124	 &	10 \\
382	&	2456592.582836	&	0.000698	 &	6 \\
\hline                                   
\end{tabular}

References: (1) Lorenz E. R. (TRESCA); (2) Naves R. (TRESCA); (3) Saral G. (TRESCA); (4) Curtis I. (TRESCA); (5) Makely N., Pree C. D. (TRESCA); (6) This work; (7) Gillier Ch. (TRESCA); (8) Shadic S. (TRESCA); (9) Sauer T. (TRESCA); (10) Pouzenc C. (TRESCA).
\end{table}

\end{document}